\documentclass[aps,showpacs,twocolumn]{revtex4}
\usepackage{amsmath}
\usepackage{graphicx}
\def\ra{\rangle}
\def\la{\langle}
\def\C{\rm C}

\begin{document}
\bibliographystyle{apsrev}

%\flushleft{\fbox{LA-UR-07-0577}}
%\hfill \fbox{\parbox[t]{1.12in}{LA-UR-07-0577}}\hspace*{0.35in}

\title{Sensitivity of Nuclear Transition Frequencies to Temporal Variation of the Fine Structure Constant or the Strong Interaction}
\author{ A.C. Hayes and J.L. Friar}
\affiliation{Theoretical Division, Los Alamos National Laboratory, 
Los Alamos, New Mexico 87545}
\date{\today}

\begin{abstract}

%%%%%%%%%%%%%%% LA-UR Rotated Box %%%%%%%%%%%%%%
\hspace*{-1.3in}
\rotatebox{90}{%
\fbox{\parbox[t]{0.88in}{LA-UR-07-0577}}
}
\vspace*{-1.05in}
%%%%%%%%%%%%%%%%%%%%%%%%%%%%%%%%%%%%%%%%

There exist in nature  a few nuclear isomers with very low (eV) excitation energies, and the combination of low energy and narrow width makes them possible candidates for laser-based investigations. The best candidate is the lowest-energy excited state known in nuclear physics, the 7.6(5) eV isomer of $^{229}$Th. A recent study suggests that a measurement  
of the temporal variation of the excitation energy of this isomer would have 5-6 orders of magnitude enhanced sensitivity to a  
variation of the fine structure constant ($\alpha \cong 1/137.036$) or of a strong interaction parameter ($m_q/\Lambda_{QCD}$). We reexamine the physics involved in these arguments. By invoking  the 
Feynman-Hellmann Theorem we argue that there is  no expectation of significantly enhanced sensitivity to  a variation in  the fine structure constant (beyond that obtained from experimental considerations such as the low energy and narrow width of the isomer).
A similar  argument applies to the strong interaction, but evaluating the shift due to temporal variations of the underlying parameters
of the strong interaction may be beyond current nuclear structure techniques.
\end{abstract}
\maketitle

The excitation energy of the first excited state of $^{229}$Th has recently been  determined\cite{becker} to be 7.6$\pm$0.5 eV. This is the lowest-lying excited state known in nuclear physics. Its lifetime is not known, but has been estimated to be roughly 5 hours\cite{becker}. These properties suggest that the transition between this state and the ground state of $^{229}$Th is within the range of  laser sources (in the vacuum ultraviolet) and that properties of the state could be probed via  high-resolution laser spectroscopy.  This possibility has led to the intriguing idea\cite{ff} that the transition could be used to place stringent constraints on a possible temporal variation of the fine structure constant ($\dot{\alpha}$) or on a temporal variation of a particular strong interaction parameter ($m_q/\Lambda_{QCD}$).  Flambaum\cite{ff} estimates that a temporal variation in the energy of this transition would provide 5-6 orders of magnitude enhanced sensitivity to a temporal variation in  these parameters. The purpose of this letter is  to examine the physics issues involved in temporal changes of nuclear transition frequencies and to quantify the corresponding sensitivity  to  variations of the underlying fundamental interactions. 

The energy of a transition ($\omega$) between  the ground state ($|g.s. \ra$) and an excited state  ($ | f \ra$) of a nucleus  is  determined by the difference of matrix elements of the nuclear Hamiltonian, $H$: 
\begin{equation}
\omega \equiv  \la f \mid H \mid f \ra - \la  g.s.\mid H  \mid g.s. \ra  \equiv \la\la H \ra\ra \, ,
\end{equation}
where we have introduced the notation $\la \la O \ra \ra$ to denote the difference of the matrix elements of the operator $O$ between the two states. Any variation in time of the transition frequency will depend on the time variation of various dynamical constants (denoted generically by $\lambda_i (t)$) embedded in the Hamiltonian, $H$. Assuming a sufficiently slow time variation in the constants, $\lambda_i$, that stationary states are well defined, using the chain rule on  Eqn.~(1) leads immediately to
\begin{equation}
\dot{\omega} = \sum_i \: \frac{\! \partial \omega}{\partial \lambda_i}\: \dot{\lambda}_i   \, .
\end{equation}

The parameter derivatives of the transition frequency can be obtained using the Feynman-Hellmann Theorem\cite{f-t,h-t}, which is sometimes (and appropriately) called the parameter theorem.  The usual statement of the theorem is that if a Hamiltonian $H(\lambda)$ depends on a parameter $\lambda$, the energy shift $E(\lambda)$ for {\it any} (bound) eigenstate of  that Hamiltonian satisfies\cite{derive}
\begin{equation}
{\partial E(\lambda) \over \partial \lambda} = \langle \Psi (\lambda) \mid \frac{\partial H(\lambda) }{\!\!\!\!\!\!\!\!\!\!\partial \lambda} \mid \Psi (\lambda) \rangle\, .
\end{equation}
Since this relation holds for the energy of every (bound) eigenstate, it must hold for the difference of two such energies. This immediately leads 
to a very simple, useful, and essentially exact form of Eqn.~(2) that applies to {\it any} system
\begin{equation}
\dot{\omega} =  \sum_i \: \left (\la f \mid \frac{\partial H}{\partial \lambda_i} \mid f \ra - \la  g.s.\mid \frac{\partial H}{\partial \lambda_i}  \mid g.s. \ra \right ) \:  \dot{\lambda}_i  \, ,
\end{equation}
or equivalently
\begin{equation}
\dot{\omega} =   \sum_i \: \left \langle \left \langle \frac{\partial H}{\partial \lambda_i} \right \rangle \right \rangle \:  \dot{\lambda}_i  \, .
\end{equation}

For the present problem we  split the nuclear Hamiltonian into  $H_n$ plus $V_{\C}$ as a convenient shorthand notation to represent respectively the strong interaction (plus kinetic energy) and the Coulomb interaction between nucleons in the nucleus. 
Let us first  restrict ourselves to a possible variation only of  the fine structure constant.  Using the fact that the Coulomb potential is linear in that constant, we find the very simple result\cite{dyson}
\begin{equation}
\dot{\omega} = \la \la V_{\C} \ra \ra \frac{\dot{\alpha}}{\alpha} =  \left ( \la f \mid V_{\C} \mid f \ra - \la  g.s.\mid V_{\C}  \mid g.s. \ra \right ) \, \frac{\dot{\alpha}}{\alpha}\, ,
\end{equation}
where the variation due to the fine structure constant is driven solely by the Coulomb energy difference\cite{note} of the two states.
Forming a fractional frequency shift\cite{steve} we have
\begin{equation}
\frac{\dot{\omega}}{\omega} = \left [\frac{\: \la \la V_{\C} \ra \ra}{\omega} \right ] \:  \frac{\dot{\alpha}}{\alpha} \, ,
\end{equation}
where the quantity in brackets can be interpreted as an enhancement factor (if larger than 1).

The conventional description of the two states in $^{229}$Th \cite{229} uses the Nilsson model\cite{nilsson}. In this picture
the  states  are described as an ``active'' neutron  in a deformed well.
In the present work we  diagonalized the 3-D Nilsson Hamiltonian for the active neutron using the deformation parameters of  M\"oller
and Nix \cite{moller-nix} to obtain wave functions for the ground state and the isomer of $^{229}$Th. 
In this model the Coulomb energies of the two states are  predicted to be the same
because the isomer and ground state are simply different Nilsson states arising from
the {\it same} deformed nuclear shape. In particular, the proton structure of the two states is identical.
The Nilsson model {\it per se} does not incorporate a Coulomb interaction, although it is implicitly part of the phenomenological interaction constants for the active nucleon\cite{nilsson,ring}. Thus, in this  model $\la\la V_{\C} \ra\ra = 0$
and there is {\it no} sensitivity to $\dot{\alpha}$. 

The Nilsson Hamiltonian does not predict these two levels of $^{229}$Th to be 7.6 eV apart 
and it is necessary to include the first order correction
to the Nilsson Hamiltonian (namely the pairing interaction) in order 
to explain the observed splitting of the states in $^{229}$Th. When 
the pairing interaction is added it plays a crucial role in reducing 
the excitation energy of the isomer of $^{229}$Th. However, 
 $\la\la V_{\C} \ra\ra$ remains zero because the proton structure of the two states remains the same\cite{ring,moller}.
 An improved estimate for $\la\la V_{\C}\ra\ra$
could be obtained by allowing the proton deformation of the two states to be different. Such a difference would arise
if different deformation parameters are needed to minimize the ground state and the isomer energies, (the M\"oller and Nix
deformation parameters were chosen to minimize the ground state energy only).
Alternatively, new proton-proton or proton-neutron terms could be added to the Hamiltonian that allow differences in the proton core for
the two states. The first of these extensions of the Nilsson plus pairing model 
is relatively straightforward to implement\cite{hayes-moller}, while the second  would require a considerably more sophisticated treatment of deformed heavy nuclei than that normally included in Nilsson-like models.
The Coulomb energy difference will likely not vanish for the real nucleus and, if experimentally possible,
the issue will best be resolved by determining the electric quadrupole moments of the two states. 
However,    
there is presently  no reason to suggest that 
$\la\la V_{\C} \ra\ra$ is dramatically larger than $\omega$,
and in any case the constraint
$\la\la H_n +V_{\C}\ra\ra = \omega$ must be satisfied.

Extracting the dependence of $\dot{\omega}$ on fundamental constants in the strong nuclear Hamiltonian $H_n$ is considerably more complicated.
It requires a sufficiently detailed understanding of that force that the appropriate parameters can be identified, and the dependence of the Hamiltonian on those parameters  must be known in order to use Eqn.~(3). 
A parameterization in terms of the quark mass $m_q$ and $\Lambda_{QCD}$ is particularly non-trivial.
An alternate  approach is to consider a parameterization of the Hamiltonian in terms of the mass of  the nucleon, the pion, etc.  
The simplest parameter relationship is then the dependence of the Hamiltonian on the nucleon mass, $M_N$.
If we ignore corrections to the nuclear potential that are inversely proportional to powers of the nucleon mass\cite{recoil}, all of that dependence is situated in the kinetic energy\cite{mass}, $T$. Performing the necessary parameter derivatives, one finds
\begin{equation}
\frac{\dot{\omega}}{\omega} = - \left [\frac{\: \la \la T \ra \ra}{\omega} \right ] \:  \frac{\dot{M}_N}{M_N} + 
 \left \langle \left \langle \frac{\partial H_n}{\partial m_\pi} \right \rangle \right \rangle \:  \frac{\dot{m_\pi}}{\omega} +  \cdots \, .
\end{equation}
Here the ellipsis denotes the dependence on all other parameters in the nuclear Hamiltonian.
Unfortunately, there  presently does not exist a model for the structure of heavy nuclei that would allow an accurate determination
 of any of these terms, with the possible
exception of the kinetic energy term.  
Within the context of the Nilsson model the kinetic energy varies from state to state, and the scale of this variation is a few MeV.

Flambaum \cite{ff} assumed  a  simple model for the structure of the $5/2^+$ ground state and the $3/2^+$ isomer of $^{229}$Th 
and   described the states in terms of the 
asymptotic Nilsson orbits $J^p[Nn_z\Lambda] =5/2^+[633]$ and $3/2^+[631]$.  In this model $\la\la T\ra\ra =0$ because 
diagonal matrix elements of the kinetic operator only depend on the quantum numbers  $N$ and $n_z$ and not on $\Lambda$. 
Hence there would be no sensitivity to
variations in the nucleon mass in this model.  However, as discussed above, the structure of 
$^{229}$Th is considerably more complicated. 
Using our wave functions from the diagonalization of the  Nilsson Hamiltonian with the deformation
parameters obtained by M\"oller and Nix\cite{moller-nix}, we find\cite{hayes-moller} that  $\la\la T\ra\ra = -0.49 $ MeV.
In this model the asymptotic states [633] and [631] 
make up about 45\%
of the wave functions for the ground state and the isomer, respectively.  
The other important orbits contributing to the wave functions are [642], [613], [622], [602] and [853] for the ground state,  and
[642], [611], [651], [622], and [871] for the isomer.  
We find that small changes in the deformation parameters from those obtained by M\"oller and Nix lead to significant changes
in the predicted structure of the two states\cite{hayes-moller} because at these deformation values several ``active'' neutron Nilsson levels 
lie very close to one another in energy. These changes lead to a correspondingly large change in $\la\la T\ra\ra$, and $\la\la T\ra\ra$ is predicted
to lie anywhere from zero to a few MeV. Thus, even the first term in Eqn.~(8) is very difficult to  predict accurately.
  
Reference \cite{ff} also investigated variation of the strong force due to variations of the quark masses, which chiral perturbation theory demonstrates is the chiral-symmetry-breaking part of the strong force. Although the strong force is not proportional to $m_q$ in leading-order, pion-mass terms (which depend on $m_q$) are incorporated into the one-pion-exchange potential, which is a part of the leading-order nuclear force contribution. Chiral perturbation theory\cite{CPT} provides enough information about the nuclear force that estimates could be made of appropriate parameter derivatives in Eqn.~(3). However, 
an accurate evaluation of Eqn.~(4) for such terms is very likely beyond the accuracy of present nuclear structure techniques.
Nevertheless, there are no obvious reasons to expect any enhanced sensitivity to temporal variations of the strong interaction parameters in $\dot{\omega}$, particularly within the framework of the Nilsson model.

Finally we comment on the origin of the large enhancement factor in Ref.\cite{ff}. 
Similar expressions to ours were obtained for the relationship  between $\dot{\omega}$ and $\dot{\alpha}$ (the strong interaction case is much more problematic), except that a parameter from the Nilsson model (whose value was approximately 1 MeV)  occurs instead of  $\la \la V_{\C} \ra \ra$. There is no justification for this. Indeed, if we use the asymptotic Nilsson model invoked in \cite{ff} or the full Nilsson Hamiltonian plus the pairing interaction, all sensitivity to $\dot{\alpha}$ drops out.

In conclusion the existence of a huge enhancement factor in the fractional frequency shift between the isomer and ground state of $^{229}$Th due to a time-dependent fine-structure constant is very difficult to realize in
any reasonable theory of the nucleus. We have examined two models for the structure of the ground state and of the isomer
of $^{232}$Th, namely, the asymptotic Nilsson model and the full Nilsson Hamiltonian using the deformation parameters
of M\"oller and Nix. In both cases there is no sensitivity to $\dot{\alpha}$ and  an extension of the Nilsson model to include 
pairing leads to the same result.
 The main issue is that the magnitude of the enhancement is strongly constrained by the Feynman-Hellmann Theorem. It would require two states of nearly identical proton structure to have  Coulomb energies that differ by $\sim$ 1 MeV. 
An analogous difficulty arises for any  enhanced sensitivity to variations of the strong interaction parameters.
The two variations (asymptotic and full) of the Nilsson model  predict $\la\la T \ra\ra$ to be zero and -0.49 MeV, respectively.
Accurately estimating a frequency shift due to variations of strong-interaction parameters, moreover, may be beyond current nuclear structure techniques. Of course, the fact that $\omega$ is very small and that the lifetime of the isomer implies a very narrow state would still be important considerations in selecting the 7.6 eV isomer of $^{229}$Th for laser-based studies.

We note in closing that many of the arguments presented here do not rely on details of nuclear structure. They are based solely on the Feynman-Hellmann Theorem and thus are equally applicable to studies
of the sensitivity of transition frequencies in atoms or other systems to temporal variations of the underlying parameters of the Hamiltonian (such as $\alpha$ or the electron mass).

\begin{acknowledgements}
We would like to thank  Gerard Jungman, Eric Lynn, Peter Milonni, Peter M\"oller, Arnold J. Sierk, Bob Wiringa and Eddy Timmermans for very helpful discussions.

\end{acknowledgements}


\begin{thebibliography}{99}
\bibitem{derive} The derivation is trivial, while the result is not obvious. We note that the derivatives of the wave functions  cancel.
\bibitem{steve} S.K. Lamoreaux (Private Communication) pointed out to us that the fractional frequency shift is not interesting in our context; only the frequency shift itself is relevant to experiments.
\bibitem{recoil} Because nuclei are weakly bound systems, one has $\la V \ra \sim - \la T \ra$. Moreover, the dominant parts of the potential are static (i.e., they do not vanish when the nucleon mass becomes very large). In other words the potential has the general form $V= V_0 + V_1/M_N +V_2/M_N^2 + \cdots$, and only $V_0$ survives in the static limit. The terms, $V_1, V_2, \cdots$, are small because the nucleon mass is so much larger than other nuclear energy scales, and therefore $V_0$ dominates. Consequently, $T$ and $V_0$ are comparable in magnitude and the other terms are significantly smaller.
\end{thebibliography}
\end{document}